\documentclass[12pt]{article}
\usepackage{epsf}
\title{ {\bf
The effects of non-universal extra dimensions on the fermion
electric dipole moments in the two Higgs doublet model.}}
\author{\vspace{1cm}\\
        {\bf E. O. Iltan}
        \thanks{E-mail address:
        eiltan@heraklit.physics.metu.edu.tr}
 \\
        Physics Department, Middle East Technical University \\
        Ankara, Turkey\\}

\date{}

\begin{document}
\setlength{\baselineskip}{24pt}
\maketitle
\setlength{\baselineskip}{7mm}
\begin{abstract}
We study the the effects of non-universal extra dimensions on the
electric dipole moments of fermions in the two Higgs doublet
model. We observe that the $t$ quark and $b$ quark electric dipole
moments are sensitive to the extra dimensions, however, in the
case of charged lepton electric dipole moments, this sensitivity
is relatively weak.
\end{abstract}
\thispagestyle{empty}
\newpage
\setcounter{page}{1}
%%%
%%%
%%
\section{Introduction}
The electric dipole moments (EDMs) of fermions are worthwhile to
study since their origin is the CP violating interaction, which is
weak in the standard model and it pushes one to investigate new
models beyond. There are number of experimental results on the
fermion EDMs in the literature. The electron, muon and tau EDMs
have been measured experimentally as $d_e =(1.8\pm 1.2\pm
1.0)\times 10^{-27} e\, cm$ \cite{Commins}, $d_{\mu} =(3.7\pm
3.4)\times 10^{-19} e\, cm$ \cite{Bailey} and $d_{\tau}
=(3.1)\times 10^{-16} e\, cm$ \cite{Groom} respectively and the
experimental upper bound of neutron EDM has been found as $d_N <
1.1\times 10^{-25} e\, cm$ \cite{Smith1}.

There is an extensive theoretical work done on the EDMs of
fermions. The calculations of quark EDMs in the framework of the
standard model (SM) \cite{sahab1} has shown that the non-zero
contribution existed at the three loop level \cite{Khiplovich} and
they were estimated as $\sim 10^{-30}\, (e-cm)$, which is a
negligible quantity. In this case, the complex phase, which is the
source of CP violation, is coming from the complex
Cabbibo-Kobayashi-Maskawa (CKM) matrix elements. Since the
numerical results of fermion EDMs are tiny in the SM, there is
enough reason to analyze these physical quantities in the
framework of new physics beyond. There are many sources of CP
violation in the models beyond the SM, such as multi Higgs doublet
models (MHDM), supersymmetric model (SUSY), extra dimensions
\cite{Schmidt},..., etc.

The EDMs of quarks was calculated in the multi Higgs doublet
models \cite{Weinberg, eril3,Iltant}, including the two Higgs
doublet model (2HDM). In these calculations $b$-quark and
$t$-quark EDMs were obtained as $10^{-21}-10^{-20} e-cm$ and
$10^{-20} e-cm$. In \cite{liao}, it was observed that the new
contributions due to the $H^{\pm}$ particles vanished at the two
loop order with the assumption that the CP violating effects came
from only the CKM matrix elements and $H^{\pm}$ particles also
mediated CP violation besides $W^{\pm}$ bosons. In \cite{Dumm}, it
was concluded that the enhancement of three orders of magnitude in
the electric dipole form factor of the $b$ quark with respect to
the prediction of 2HDM I and II was possible. \cite{Xu} is devoted
to the calculation of leading contribution to the EDM of the top
quark in Higgs-boson-exchange models of CP nonconservation and the
dipole moments were estimated of the order $10^{-20}\, (e-cm)$. In
\cite {Bhaskar} lepton electric dipole moments in the
supersymmetric seesaw model has been studied. In \cite
{Iltmuegam}, the EDM of the electron has been predicted as $d_e$
as $10^{-32}\, e-cm$ using the experimental result of $d_{\mu}$
and the upper limit of $BR(\mu\rightarrow e\gamma)$. The work
\cite{IltanNonCom} is devoted to quark and lepton EDM moments in
the framework of the SM with the inclusion of non-commutative
geometry. In recent works, the EDMs of nuclei, deutron, neutron
and some atoms have been studied extensively \cite{Vladimir}.

Our work is devoted to the investigation of the fermion EDMs in
the case that the CP violating interactions are carried by complex
Yukawa couplings appearing in the flavor changing (FC) neutral
current vertices, in the model III version of the 2HDM, with the
inclusion of the extra dimensions. There is an extensive work on
the extra dimensions in the literature
\cite{Appelquist}-\cite{Lam}. The main motivation of such
dimensions is to find a solution to the gauge hierarchy problem of
the SM. The effects of each extra dimension is felt with the
production of Kaluze-Klein (KK) states of the fields, which are
obtained after the compactification on a circle of radius R. The
number $1/R$ is known as the compactification scale and the its
size have been estimated in the range $200-500\, GeV$, using
electroweak precision measurements \cite{Appelquist}, the \(
B-\bar{B} \) -mixing \cite{Papavassiliou}, \cite{Chakraverty} and
the flavor changing process $b \to s \gamma$ \cite{Agashe}.
Furthermore, this size has been obtained as large as few hundereds
of GeV in several works
\cite{Arkani,Antoniadis1,Antoniadis2,Carone,Antoniadis3}. If all
the fields live in higher dimensions \cite{Appelquist,Carone},
such extra dimensions are called as 'universal extra dimensions'
(UED's), and in this case the extra dimensional momentum, and
therefore the KK number at each vertex, is conserved. As a result
of KK number conservation, the KK modes enter into the
calculations as loop corrections. However, if some of the
particles do not feel the extra dimensions and others do, namely
non-universal extra dimension case, the coupling of two zero modes
with the KK mode is switched on and  the contributions of extra
dimensions to the tree level processes become non-zero.

In the present work, we consider the effects of non-universal
extra dimensions on the EDMs of fermions by assuming that the new
Higgs doublet and the gauge sector feel the extra dimensions,
however, the other SM fields do not feel and are confined on 4D
brane. Notice that in the case of UED, where all the fields are
accessible to the extra dimensions, there does not exist any new
contribution to the fermion EDMs due to the KK modes of Higgs
fields and fermions, at least in the one loop level. The higher
dimensional effects on the EDMs of fermions are carried by the
intermediate charged, $H^\pm$, and neutral Higgs, $h^0$ and $A^0$,
fields with the vertices including "two zero modes-KK mode". We
study those additional effects for one and two spatial extra
dimensions.

In the numerical calculations, we observe that the $t$ quark and
$b$ quark EDMs are sensitive to the extra dimension, especially
the double one, however, in the case of charged lepton EDMs, this
sensitivity is relatively weak. Therefore, the future accurate
measurements of EDMs of fermions may be an effective tool to check
the existence and the number of extra dimensions and the
restriction of the compactification scale.

The paper is organized as follows: In Section 2, we present EDMs
of fermions, t-quark, b-quark and charged leptons, in the model
III version of the 2HDM with the inclusion of the non-universal
extra dimensions. Section 3 is devoted to discussion and our
conclusions.
%%%
%%%
\section{Electric dipole moments of fermions in the two  Higgs
doublet model with the inclusion of extra dimensions}
The existence of the fermion EDM depends on the CP violating
fermion-fermion-photon interaction. In the framework of the SM,
the CP violation is carried by the complexity of the Cabbibo
Cobayashi Maskawa (CKM) matrix elements and the estimated
numerical values of EDMs of fermions are extremely small. This
makes it charming to investigate new complex phases by considering
the physics beyond the SM. The model III version of the 2HDM is
one of the candidate since the FC neutral currents (FCNC) are
permitted at tree level and the new Yukawa couplings can be
complex in general. With the addition of spatial extra dimensions
which are felt by the new Higgs doublet there appear additional
contributions sensitive to the compactification scale $1/R$ where
$R$ is the radius of the compactification. The Yukawa Lagrangian
responsible for the EDM fermion in such a single extra dimension
reads:
\begin{eqnarray}
{\cal{L}}_{Y}&=&\eta^{U}_{ij} \bar{Q}_{i L} \tilde{\phi_{1}} U_{j
R}+ \eta^{D}_{ij} \bar{Q}_{i L} \phi_{1} D_{j R}+
\xi^{U\,\dagger}_{5 \,ij} \bar{Q}_{i L} (\tilde{\phi_{2}}|_{y=0})
U_{j R}+ \xi^{D}_{5 \,ij} \bar{Q}_{i L} (\phi_{2}|_{y=0}) D_{j R}
\nonumber \\
&+& \eta^{E}_{ij} \bar{l}_{i L} \phi_{1} E_{j R}+ \xi^{E}_{5 \,ij}
\bar{l}_{i L} (\phi_{2}|_{y=0}) E_{j R} + h.c. \,\,\, ,
\label{lagrangian}
\end{eqnarray}
where $y$ represents the extra dimension, $L$ and $R$ denote
chiral projections $L(R)=1/2(1\mp \gamma_5)$, $\phi_{i}$ for
$i=1,2$, are the two scalar doublets, $\bar{Q}_{i L}$ are left
handed quark doublets, $U_{j R} (D_{j R})$ are  right handed up
(down) quark singlets, $l_{i L}$ ($E_{j R}$) are lepton doublets
(singlets), with family indices $i,j$. The Yukawa couplings
$\xi^{E,D,U}_{5\, ij}$, complex in general, are dimensionful and
rescaled to the ones in 4-dimension as $\xi^{U,D,E}_{5\,
ij}=\sqrt{2 \pi R}\,\, \xi^{U,D,E}_{ij}$
\footnote{In the following we use the dimensionful coupling
$\bar{\xi}^{U,D,E}_{N}$ with the definition
$\xi^{U,D,E}_{N,ij}=\sqrt{\frac{4\, G_F}{\sqrt{2}}}\,
\bar{\xi}^{U,D,E}_{N,ij}$ where N denotes the word "neutral".}
Here we choose the Higgs doublets $\phi_{1}$ and $\phi_{2}$ as
\begin{eqnarray}
\phi_{1}=\frac{1}{\sqrt{2}}\left[\left(\begin{array}{c c}
0\\v+H^{0}\end{array}\right)\; + \left(\begin{array}{c c}
\sqrt{2} \chi^{+}\\ i \chi^{0}\end{array}\right) \right]\, ;
\phi_{2}=\frac{1}{\sqrt{2}}\left(\begin{array}{c c}
\sqrt{2} H^{+}\\ H_1+i H_2 \end{array}\right) \,\, .
\label{choice}
\end{eqnarray}
with the vacuum expectation values,
\begin{eqnarray}
<\phi_{1}>=\frac{1}{\sqrt{2}}\left(\begin{array}{c c}
0\\v\end{array}\right) \,  \, ;
<\phi_{2}>=0 \,\, .
\label{choice2}
\end{eqnarray}
and collect SM (new) particles in the first (second) doublet.
Notice that $H_1$ and $H_2$ are the mass eigenstates $h^0$ and
$A^0$ respectively since no mixing occurs between two CP-even
neutral bosons $H^0$ and $h^0$ at tree level in our case.

Since the new Higgs field $\phi_{2}$ is accessible to extra
dimension, the compactification on a circle of radius $R$ results
in the expansion of $\phi_{2}$ into its KK modes as
\begin{eqnarray}
\phi_{2}(x,y ) & = & {1 \over {\sqrt{2 \pi R}}} \left\{
\phi_{2}^{(0)}(x) + \sqrt{2} \sum_{n=1}^{\infty} \phi_{2}^{(n)}(x)
\cos(ny/R)\right\} \, \label{SecHiggsField}
\end{eqnarray}
where $\phi_{2}^{(0)}(x)$ the 4-dimensional Higgs doublet which
contains the charged Higgs boson $H^+$, the neutral CP even (odd)
$H^1$ ($H^2$) Higgs bosons. The non-zero KK mode of Higgs doublet
$\phi_{2}$ includes a charged Higgs of mass
$\sqrt{m_{H^\pm}^2+m_n^2}$, a neutral CP even Higgs of mass
$\sqrt{m_{h^0}^2+m_n^2}$, a neutral CP odd Higgs of mass
$\sqrt{m_{A^0}^2+m_n^2}$ where $m_n=n/R$ is the mass of $n$'th
level KK particle. In addition to the new Higgs field the gauge
fields feel the extra dimensions and therefore there exist their
KK modes after the compactification, however they do not bring new
contributions to the EDM of fermions.

\begin{figure}[htb]
\vskip 1.0truein \centering \epsfxsize=6.4in
\leavevmode\epsffile{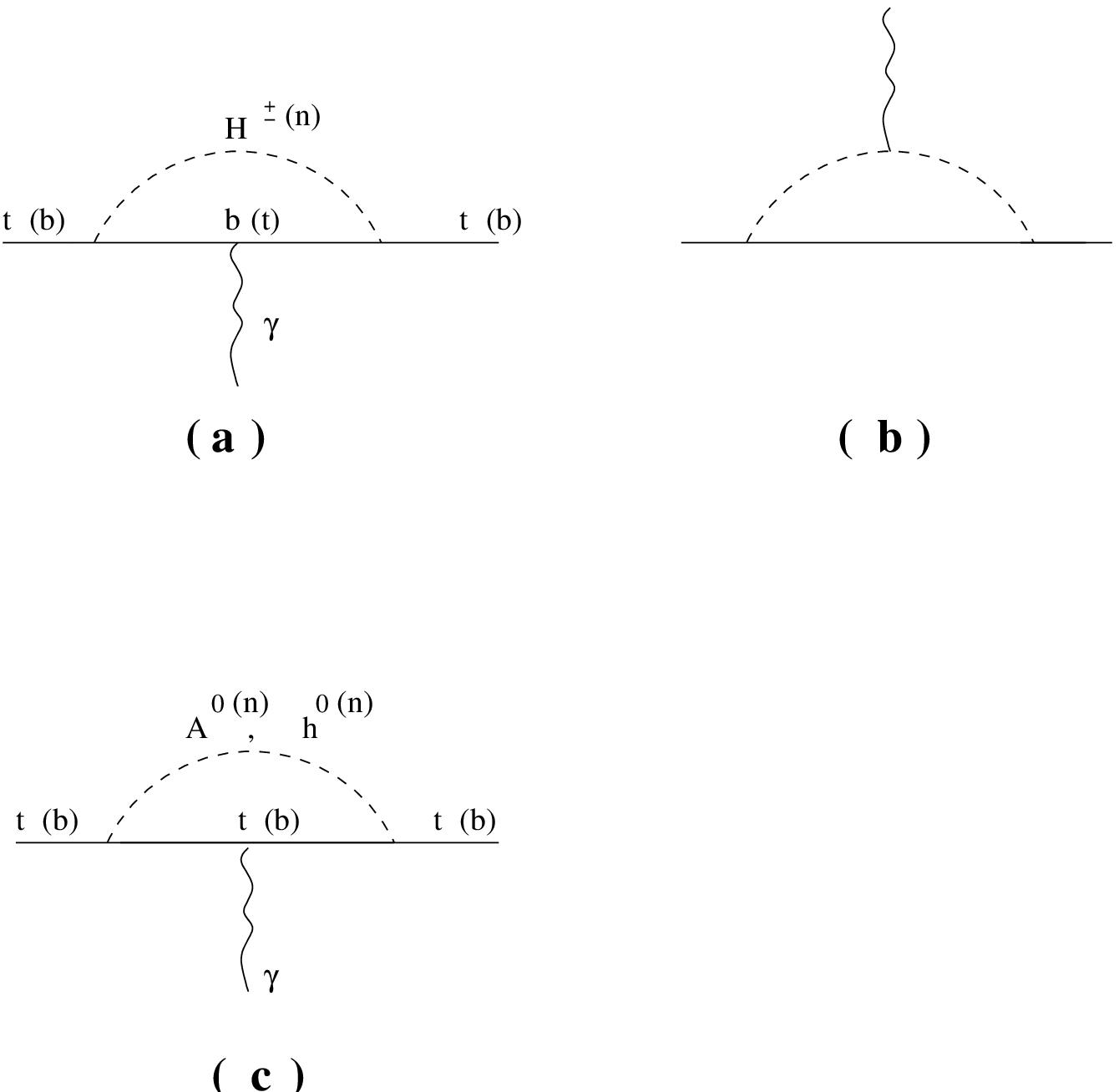} \vskip 1.0truein \caption[]{One loop
diagrams contribute to EDM of top and bottom quarks due to charged
Higgs boson $H^{\pm}$ (\textbf{a} and \textbf{b}) and neutral
Higgs bosons $h^0$, $A^0$ (\textbf{c}) in the 2HDM, including KK
modes. Wavy lines represent the electromagnetic field and dashed
lines the Higgs field.} \label{fig1}
\end{figure}

\begin{figure}[htb]
\vskip 0.5truein \centering \epsfxsize=2.8in
\leavevmode\epsffile{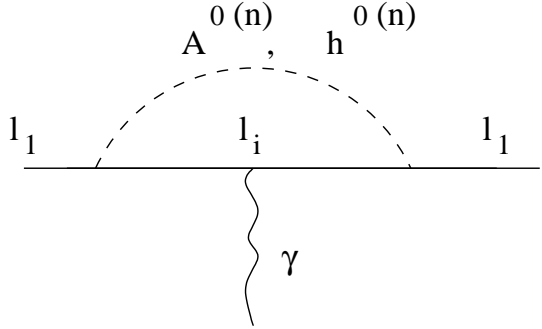} \vskip 0.5truein \caption[]{One loop
diagrams contribute to EDM of charged leptons due to neutral Higgs
bosons $h^0$, $A^0$ in the 2HDM, including KK modes. Wavy lines
represent the electromagnetic field and dashed lines the Higgs
field where $l_{1 \,(i)}=e, \mu, \tau$} \label{fig2}
\end{figure}

The effective EDM interaction for a fermion $f$ is given by
\begin{eqnarray}
{\cal L}_{EDM}=i e d_f \,\bar{f}\,\gamma_5 \,\sigma^{\mu\nu}\,f\,
F_{\mu\nu} \,\, , \label{EDM1}
\end{eqnarray}
where $F_{\mu\nu}$ is the electromagnetic field tensor, '$d_{f}$'
is EDM of the fermion and it is a real number by hermiticity. In
Figs. \ref{fig1} and \ref{fig2} we present the 1-loop diagrams
which contribute to the EDMs of fermions with the help of the
complex Yukawa couplings. Here, the charged Higgs contributions
play the important role in the case of the quarks. However, for
charged leptons we assume that there is no CKM type lepton mixing
matrix and, therefore, only the neutral Higgs part gives a
contribution to their EDMs.

Now, we would like to present the EDMs of fermions ($t$ quark, $b$
quark and charged leptons) with the addition of a single
non-universal extra dimension where the second Higgs doublet and
gauge sector feels the extra dimensions and all other SM particles
are restricted to the 4D brane. The top quark EDM  \cite{Iltant}
reads:
\begin{eqnarray}
d_t=d^0_t+2\, \sum_{n=1}^{\infty} \Bigg( (\int_{0}^{1} dx \,
d_{t}^{H^{\pm}})+d_{t}^{h^0}+d_{t}^{A^0} \Bigg)\,\,, \label{EDMt}
\end{eqnarray}
where
\begin{eqnarray}
d_t^{H^{\pm}}&=&\frac{4\,G_F}{\sqrt {2}}\frac{1}{48\pi^2}\,
\frac{m_b}{m_t^2}\,Im(\bar{\xi}^{D\,*}_{N,bb}\,\bar{\xi}^{U}_{N,tt})\,
|V_{tb}|^2 \, \frac{(-1\,+\,x)\,\big(Q_b\, (-1\,+\,x)+3\, x\, \big
) \,y_t} {r_b\, y_t+x^2\, y_t-x\,(-1+y_t+r_b\,y_t)}
\,\, , \nonumber \\
d_t^{h^0}&=&\frac{4\,G_F}{\sqrt {2}}\frac{1}{16\pi^2}\,
\frac{1}{m_t}\,Im(\bar{\xi}^{U}_{N,tt})\,Re(\bar{\xi}^{U}_{N,tt})
\, Q_t\, \Bigg\{ 1-\frac{r_1}{2}\,ln\,r_1 \nonumber \\
&+& \frac{r_1\,(r_1-2)}{\sqrt{r_1\,(r_1-4)}}\Big ( Arctan\, \big(
\frac{r_1}{\sqrt{r_1\,(r_1-4)}}\big ) - Arctan\, \big(\frac{r_1-2}
{\sqrt{r_1\,(r_1-4)}} \big) \Big) \Bigg\} \,\, , \mbox{ for
$r_1<4$}, \nonumber \\  \nonumber \\  \nonumber\\
d_t^{h^0}&=&-\frac{4\,G_F}{\sqrt {2}}\frac{1}{16\pi^2}\,
\frac{1}{m_t}\,Im(\bar{\xi}^{U}_{N,tt})\,Re(\bar{\xi}^{U}_{N,tt})
\, Q_t \, \Bigg\{ 1-\frac{r_1\,(r_1-2)}{\sqrt{r_1\,(r_1-4)}}\,
ln\frac{\sqrt{r_1}-\sqrt{r_1-4}}{2} \nonumber \\ &-& \frac{1}{2}
r_1 \,ln\, r_1 \Bigg\} \,\, , \mbox{ for $r_1>4$}\, ,
\nonumber \\
d_t^{A^0}&=&-d_t^{h^0} (r_1\rightarrow r_2) \,\, , \label{EDMt2}
\end{eqnarray}
$d^0_t$ is the contribution without the extra dimension, namely
$n=0$ case, and $r_b=\frac{m_b^2}{m_t^2}$,
$r_1=\frac{m_{h^0}^2+n^2/R^2}{m_t^2}$,
$r_2=\frac{m_{A^0}^2+n^2/R^2}{m_t^2}$,
$y_t=\frac{m_t^2}{m^2_{H^{\pm}}+n^2/R^2}$, $Q_b$ and $Q_t$ are
charges of $b$ and $t$ quarks respectively.

In eqs. (\ref{EDMt2}), we take only internal $b\,(t)$-quark
contribution for charged (neutral) Higgs interactions. Here, we
consider that the Yukawa couplings $\bar{\xi}^{U}_{N,it},\, i=u,c
$, and $\bar{\xi}^{D}_{N, bj},\, j=d,s $ are negligible compared
to $\bar{\xi}^{U}_{N,tt}$ and $\bar{\xi}^{D}_{N,bb}$ (see
\cite{Iltmuegam}). Furthermore we use the parametrization
\begin{eqnarray}
\bar{\xi}^{U}_{N,tt}=|\bar{\xi}^{U}_{N,tt}|\, e^{i \theta_{t}}
\nonumber \, , \\
\bar{\xi}^{D}_{N,bb}=|\bar{\xi}^{D}_{N,bb}|\, e^{i \theta_{b}} \,
, \label{xi}
\end{eqnarray}
for the complex Yukawa couplings $\bar{\xi}^{U}_{N,tt}$ and
$\bar{\xi}^{D}_{N,bb}$. Notice that the neutral Higgs
contributions to the $d_t$ are switched on if we choose the
coupling $\bar{\xi}^{U}_{N,tt}$ complex. In our numerical
calculations for $d_t$ we choose this coupling complex to
determine the strength of the the neutral Higgs contributions and
observe that they are small compared to the charged Higgs ones.

Similarly the $b$ quark EDM  \cite{eril3} reads:
\begin{eqnarray}
d_b=d^0_b+2\,\sum_{n=1}^{\infty}\,
(d_b^{H^{\pm}}+d_b^{h^0}+d_b^{A^0})\,\,, \label{EDMb}
\end{eqnarray}
where $d_b^{H^{\pm}}$ , $d_b^{h^0}$ and $d_b^{A^0}$ are
\begin{eqnarray}
d_b^{H^{\pm}}\!\!\!&=& \!\!\!\!\!\!\frac{4\,G_F}{\sqrt {2}}
\frac{e}{32\pi^2}\, \frac{1}{m_t}\, \bar{\xi}^{U}_{N,tt}\,
Im(\bar{\xi}^{D}_{N,bb})\, |V_{tb}|^2 \, \frac{y_t\, \big(
(-1\,+\,Q_t\, (-3\,+\,y_t)-\,y_t)(y_t\,-\,1)+2\, (Q_t\,+\,y_t)
\,ln\,y_t \big)} {(y_t-1)^3}
\,\, , \nonumber \\
d_b^{h^0}&=&-\frac{4\,G_F}{\sqrt {2}} \frac{e}{16\pi^2}\,
\frac{Q_b}{m_b}\, Im(\bar{\xi}^{D}_{N,bb}) \,  Re
(\bar{\xi}^{D}_{N,bb})\,
(1-\frac{r_1\,(r_1-2)}{\sqrt{r_1\,(r_1-4)}}\,
ln\frac{\sqrt{r_1}-\sqrt{r_1-4}}{2}-\frac{1}{2} r_1 \,ln\, r_1 )
\,\, , \nonumber \\
d_b^{A^0}&=&-d^{h^0} (r_1\rightarrow r_2) \,\, . \label{EDMb2}
\end{eqnarray}
where $d^0_b$ is the contribution without the extra dimension. In
eq. (\ref{EDMb2}), we take into account only internal $t$-quark
contribution for charged Higgs boson and internal $b$-quark
contribution for neutral Higgs interactions by following the
previous assumption. Furthermore, we choose $\bar{\xi}^{U}_{N,tt}$
real and $\bar{\xi}^{D}_{N,bb}$ complex since the $d_b$ is
sensitive (not sensitive) to the imaginary part of the coupling
$\bar{\xi}^{D}_{N,bb}$ ($\bar{\xi}^{U}_{N,tt}$).

Finally, we will present the charged lepton EDMs with the addition
of non-universal extra dimensions. Since there is no CKM type
lepton mixing matrix according to our assumption only the neutral
Higgs part gives a contribution to their EDMs and $l$-lepton EDM
'$d_l$' $(l=e,\,\mu,\,\tau)$  can be calculated as a sum of
contributions coming from neutral Higgs bosons $h_0$ and $A_0$
\cite{Iltmuegam},
\begin{eqnarray}
d_l= -\frac{i G_F}{\sqrt{2}} \frac{e}{32\pi^2}\,
\frac{Q_{\tau}}{m_{\tau}}\, ((\bar{\xi}^{D\,*}_{N,l\tau})^2-
(\bar{\xi}^{D}_{N,\tau l})^2)\, \Bigg ( (F_1 (y_{h_0})-F_1
(y_{A_0}))+2\, \sum_{n=1}^{\infty}\, (F_1 (y^n_{h_0})-F_1
(y^n_{A_0})) \Bigg), \label{emuEDM}
\end{eqnarray}
for $l=e,\mu$ and
\begin{eqnarray}
d_{\tau}&=& -\frac{i G_F}{\sqrt{2}} \frac{e}{32\pi^2}\, \Big{\{}
\frac{Q_{\tau}}{m_{\tau}}\, ((\bar{\xi}^{D\,*}_{N,\tau\tau})^2-
(\bar{\xi}^{D}_{N,\tau \tau})^2)\, \Bigg( (F_2
(r_{h_0})-F_2(r_{A_0}))+ 2\,
\sum_{n=1}^{\infty}\, (F_2 (r^n_{h_0})-F_2(r^n_{A_0}))\Bigg) \nonumber \\
&-&  Q_{\mu}\, \frac{m_{\mu}}{m^2_{\tau}}
((\bar{\xi}^{D\,*}_{N,\mu\tau})^2-(\bar{\xi}^{D}_{N,\tau
\mu})^2)\, \Bigg((r_{h_0}\,ln\, (z_{h_0})-r_{A_0}\,ln\,
(z_{A_0}))\nonumber \\ &+& 2\, \sum_{n=1}^{\infty}\,
(r^n_{h_0}\,ln\, (z^n_{h_0})-r^n_{A_0}\,ln\, (z^n_{A_0}))\Bigg)
\Big{\}} \,\, , \label{tauEDM}
\end{eqnarray}
where the functions $F_1 (w)$, $F_2 (w)$ and $F_3 (w)$ are
\begin{eqnarray}
F_1 (w)&=&\frac{w\,(3-4\,w+w^2+2\,ln\,w)}{(-1+w)^3}\nonumber \,\, , \\
F_2 (w)&=& w\, ln\,w + \frac{2\,(-2+w)\, w\,ln\,
\frac{1}{2}(\sqrt{w}-\sqrt{w-4})}{\sqrt{w\,(w-4)}} \,\, .
\label{functions1}
\end{eqnarray}
Here $y^n_{H}=\frac{m^2_{\tau}}{m^2_{H}+n^2/R^2}$,
$r^n_{H}=\frac{1}{y^n_{H}}$ and
$z^n_{H}=\frac{m^2_{\mu}}{m^2_{H}+n^2/R^2}$, $y_{H}=y^0_{H}$,
$r_{H}=r^0_{H}$ and $z_{H}=z^0_{H}$, $Q_{\tau}$ and $Q_{\mu}$ are
charges of $\tau$ and $\mu$ leptons respectively. In eq.
(\ref{emuEDM}) we take into account only internal $\tau$-lepton
contribution respecting our assumption that the Yukawa couplings
$\bar{\xi}^{D}_{N, ij},\, i,j=e,\mu$, are small compared to
$\bar{\xi}^{D}_{N,\tau\, i}\, i=e,\mu,\tau$ due to the possible
proportionality of the Yukawa couplings to the masses of leptons
in the vertices. In eq. (\ref{tauEDM}) we present also the
internal $\mu$-lepton contribution, which can be neglected
numerically. Notice that, we make our calculations in arbitrary
photon four momentum square $q^2$ and take $q^2=0$ at the end. We
used the parametrization
\begin{eqnarray}
\bar{\xi}^{D}_{N,\tau l}=|\bar{\xi}^{D}_{N,\tau l}|\, e^{i
\theta_{l}} \,\, , \label{xi2}
\end{eqnarray}
and the Yukawa factors in eqs. (\ref{emuEDM}) and  (\ref{tauEDM})
can be written as
\begin{eqnarray}
((\bar{\xi}^{D\,*}_{N,l\tau})^2-(\bar{\xi}^{D}_{N,\tau
l})^2)=-2\,i sin\,2\theta_{l}\, |\bar{\xi}^{D}_{N,\tau l}|^2
\end{eqnarray}
where $l=e,\mu,\tau$. Here $\theta_{l }$ is CP violating
parameters which is the source of the lepton EDM.

In the case two extra spatial dimensions which are felt by the
second Higgs doublet $\phi_{2}$ the compactification of the extra
dimensions is done on a torus $S^1\times S^1$ and the doublet
$\phi_{2}$ can be expanded into its KK modes as
\begin{eqnarray}
\phi_{2}(x,y,z ) & = & {1 \over {2 \pi R}} \left\{
\phi_{2}^{(0,0)}(x) + 2 \sum_{n,r} \phi_{2}^{(n,r)}(x) \cos(n
y/R+rz/R)\right\} \, . \label{SecHiggsField2}
\end{eqnarray}
Here the indices  $n$ and $r$ are positive integers including zero
but both are not zero at the same time and each circle is
considered the same radius $R$. Furthermore, $\phi_{2}^{(0,0)}(x)$
is the 4-dimensional Higgs doublet including the charged Higgs
boson $H^\pm$, the neutral CP even (odd) $h^0$ ($A^0$) Higgs
bosons. The non-zero KK mode $\phi_{2}^{(n,r)}(x)$ ($n$ or $r$ is
different than $0$ ) of Higgs doublet $\phi_{2}$ contains  a
charged Higgs of mass $\sqrt{m_{H^\pm}^2+m_n^2+m_r^2}$, a neutral
CP even Higgs of mass $\sqrt{m_{h^0}^2+m_n^2+m_r^2}$, a neutral CP
odd Higgs of mass $\sqrt{m_{A^0}^2+m_n^2+m_r^2}$ where the mass
terms $m_n=n/R$ and $m_r=r/R$ exist due to the compactification.
%%%
%%%
\section{Discussion}
In this section we analyze the effects non-universal extra
dimensions on the EDMs of fermions, including top quark, bottom
quark and charged leptons. Since the EDM interaction is CP
violating, one needs a CP violating phase and we consider the
complex Yukawa couplings appearing in the FCNC  at tree level in
the framework of the model III.

For the top quark EDM, the CP violating parameters are
$sin\,\theta_{b}$ and $sin\,\theta_{t}$ play the main role (see
eq. (\ref{xi})). In this case, the neutral Higgs boson
contribution exists if the parameter $sin\,\theta_{t}$ is
non-zero, namely, the Yukawa coupling drives $t-t-h^0 (A^0)$
interaction has a complex phase. On the other hand the bottom
quark EDM appears with the non-zero parameter $sin\,\theta_{b}$,
which is the imaginary part of the Yukawa coupling responsible for
the $b-b-h^0 (A^0)$ interaction. These CP violating parameters,
the Yukawa couplings, $\bar{\xi}^{U (D)}_{N, ij}$, and the masses
of new Higgs bosons, $H^{\pm}$, $h^0$ and $A^0$ are the free
parameters of the model used and they should be restricted using
the experimental measurements.

In our numerical calculations we neglect all the quark sector
Yukawa couplings except $\bar{\xi}^{U}_{N, tt}$ and
$\bar{\xi}^{D}_{N, bb}$ since they are negligible due to their
light flavor contents, by our assumption, similar to the
Cheng-Sher scenario \cite{Sher}. Furthermore, for the couplings
$\bar{\xi}^{U}_{N, tt}$, $\bar{\xi}^{D}_{N, bb}$ and the CP
violating parameters $sin\,\theta_{t}$ and $sin\,\theta_{b}$, we
use the constraint region which is obtained by restricting the
Wilson coefficient $C_7^{eff}$ in the region $0.257 \leq
|C_7^{eff}| \leq 0.439$. Here upper and lower limits were
calculated using the CLEO measurement \cite{cleo2}
\begin{eqnarray}
Br (B\rightarrow X_s\gamma)= (3.15\pm 0.35\pm 0.32)\, 10^{-4} \,\, .
\label{br2}
\end{eqnarray}
and all possible uncertainities in the calculation of $C_7^{eff}$
\cite{alil1}. Notice that $C_7^{eff}$ is the effective coefficient
of the operator $O_7 = \frac{e}{16 \pi^2} \bar{s}_{\alpha}
\sigma_{\mu \nu} (m_b R + m_s L) b_{\alpha} {\cal{F}}^{\mu \nu}$
(see \cite{alil1} and references therein). We respect also the
additional constraint for the angle $\theta_{t}$ and $\theta_{b}$,
comes from the experimental upper limit of neutron electric dipole
moment, $d_n<10^{-25}\hbox{e$\cdot$cm}$, which leads to
$\frac{1}{m_t m_b} Im(\bar{\xi}^{U}_{N, tt}\, \bar{\xi}^{* D}_{N,
bb})< 1.0$ for $M_{H^\pm} \approx 200$ GeV \cite{david}. In our
numerical calculations we choose the upper limit for
$C_7^{eff}>0$, fix $\bar{\xi}^{D}_{N,bb}=30\,m_b$ and take
$\bar{\xi}^{U}_{N, tt}\sim 35\, GeV$, respecting the constraints
mentioned.

The charged lepton EDM depends on the leptonic complex Yukawa
couplings $\bar{\xi}^D_{N,ij}, i,j=e, \mu, \tau$ which are another
set of free parameters in the model III. Similar to the previous
assumptions, we consider the Yukawa couplings
$\bar{\xi}^{D}_{N,ij},\, i,j=e,\mu $, as smaller compared to
$\bar{\xi}^{D}_{N,\tau\, i}\, i=e,\mu,\tau$ and we assume that
$\bar{\xi}^{D}_{N,ij}$ is symmetric with respect to the indices
$i$ and $j$. For the coupling $\bar{\xi}^{E}_{N,\tau \mu}$ we use
the restriction (see \cite{Iltananomuon}) coming from the
experimental uncertainty, $10^{-9}$, in the measurement of the
muon anomalous magnetic moment \cite{BNL}. Notice that the
predicted upper limit for the coupling $\bar{\xi}^{E}_{N,\tau
\mu}$ is $30\, GeV$ in \cite{Iltananomuon}. For the coupling
$\bar{\xi}^{E}_{N,\tau \tau}$ there is no stringent prediction and
we take an intermediate value which is greater than the coupling
$\bar{\xi}^{E}_{N,\tau \mu}$.

The inclusion of the spatial extra dimensions that are felt by the
new Higgs doublet results in the new contributions, emerging from
the KK excitations of the new charged and neutral Higgs fields, to
the EDMs of fermions. The new vertices appearing in the
calculation of EDMs are coming from the fermion-fermion-KK Higgs
interaction where the KK number is not conserved in contrast to
the case of UED \cite{Appelquist, Carone}. The compactification of
extra dimensions to the torus brings new parameter, called the
compactification radius $R$. Here   $R$ is the size of the extra
dimension and it needs to be restricted. The lower bound for
inverse of the compactification radius is estimated as $\sim 300\,
GeV$ \cite{Appelquist}.

In our calculations we study the effects of one and two spatial
non-universal extra dimensions on the EDMs of fermions, top and
bottom quarks, charged leptons $\mu$ and $\tau$.  In the case of
two spatial extra dimensions the compactification is done on a
torus $S^1\times S^1$ and we assume that each circle has the same
radius $R$. We see that the EDMs of fermions we consider are
sensitive to the extra dimensions. Notice that we do not present
the new effect on the light fermions, other quarks and electron,
since the corresponding Yukawa couplings and the mass ratios
$m_{Light}/m_{H^{\pm}, h^0, A^0}$ are highly suppressed. In
addition to this we take the Yukawa couplings for
fermion-fermion-KK Higgs interaction same as the ones existing in
the zero-mode case with the assumption that the Yukawa couplings
are sensitive to the flavor.

\begin{figure}[htb]
\vskip -3.0truein \centering \epsfxsize=6.8in
\leavevmode\epsffile{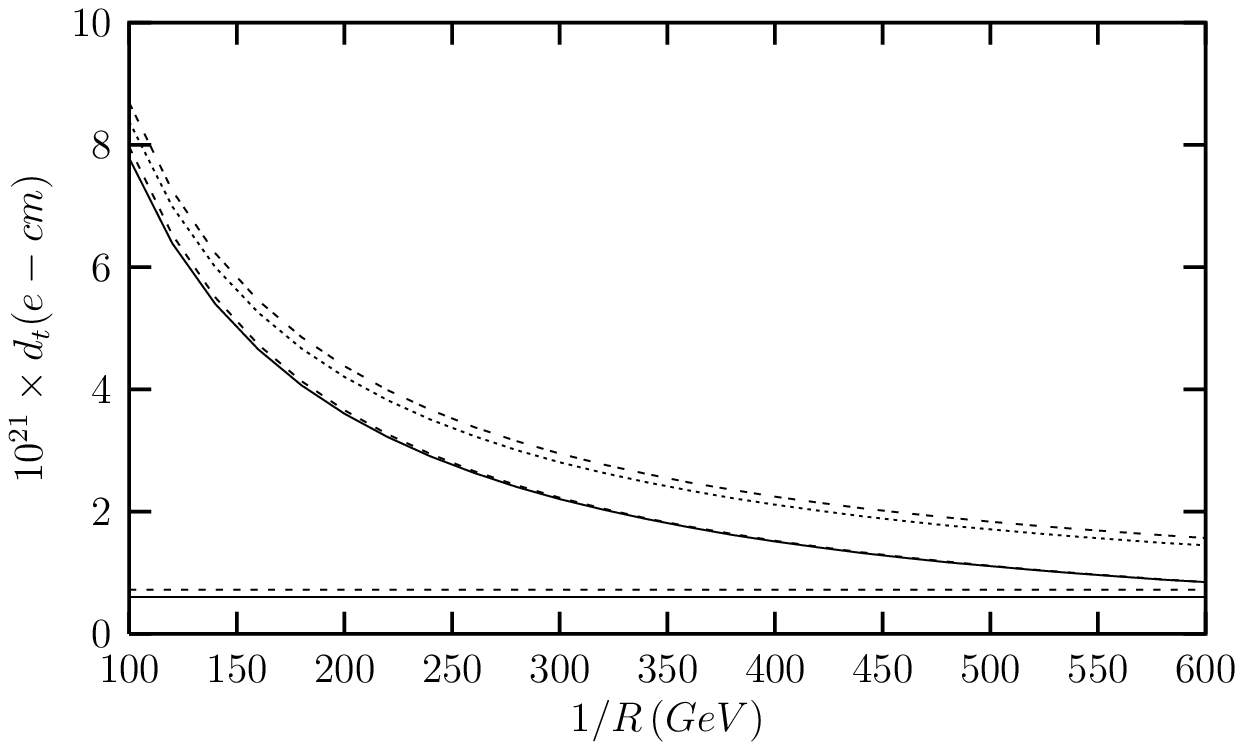} \vskip -3.0truein
\caption[]{$d_t$ with respect to compactification scale $1/R$ for
$m_{H^{\pm}}=400\, GeV$, $m_{h^0}=85\, GeV$, $m_{A^0}=90\, GeV$,
$sin\,\theta_{t}=0.1$ and $sin\,\theta_{b}=0.5$, in the case of a
single extra dimension. Here the solid (dashed) line represents
charged Higgs (total) contribution to $d_t$ without extra
dimension,  solid (dashed) curve represents charged Higgs (total)
single extra dimension contribution to $d_t$ and dotted (double
dashed) curve represents charged Higgs (total) total contribution
to the $d_t$.} \label{EDMttotExtr1R}
\end{figure}

In  Fig. \ref{EDMttotExtr1R}, we plot EDM $d_t$ with respect to
compactification scale $1/R$ for $m_{H^{\pm}}=400\, GeV$,
$m_{h^0}=85\, GeV$, $m_{A^0}=90\, GeV$, the small value of
$sin\,\theta_{t}=0.1$ and the intermediate value of
$sin\,\theta_{b}=0.5$. Here the solid (dashed) line represents
charged Higgs (total) contribution to the t-quark EDM without
extra dimension,  solid (dashed) curve represents charged Higgs
(total) contribution of a single extra dimension to the t-quark
EDM and dotted (double dashed) curve represents charged Higgs
(total) total contribution to the t-quark EDM. This figure shows
that the contribution of the extra dimensions are, especially for
the charged Higgs part, larger than the one without the extra
dimension, for $1/R\leq 600\, GeV$. In fact, in the case of
vanishing $sin\,\theta_t$ the neutral Higgs bosons $h^0$ and $A^0$
and their KK modes do not have any contribution to the EDM of top
quark and for nonzero $sin\,\theta_t$, they are suppressed even
with the addition of extra dimensions. The EDM of top quark
reaches the numerical value $3.0\times 10^{-21}\, (e-cm)$ for
$1/R\sim 300\, GeV$ with the inclusion of a single extra dimension
and it is almost six times larger than the one obtained without
extra dimensions. This is an interesting result since the EDM of
top quark is sensitive to a single extra dimension and it may
ensure a powerful information about the existence of extra
dimensions and the restriction of the compactification scale, with
the forthcoming experimental measurements.

\begin{figure}[htb]
\vskip -3.0truein \centering \epsfxsize=6.8in
\leavevmode\epsffile{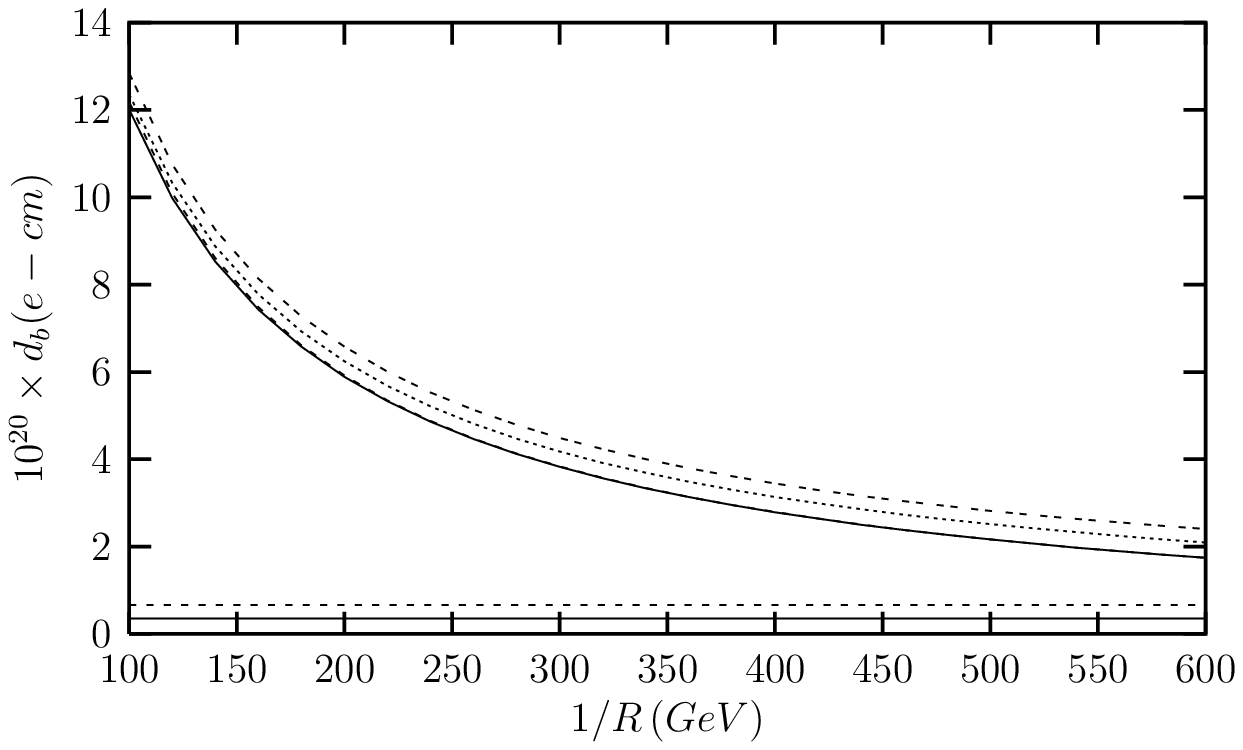} \vskip -3.0truein
\caption[]{The same as Fig. \ref{EDMttotExtr1R} but for $d_b$ and
$sin\,\theta_{b}=0.5$}. \label{EDMbtotExtr1R}
\end{figure}

Fig. \ref{EDMbtotExtr1R}, is devoted to the compactification scale
$1/R$ dependence of the EDM $d_b$  for $m_{H^{\pm}}=400\, GeV$,
$m_{h^0}=85\, GeV$, $m_{A^0}=90\, GeV$, and the intermediate value
of $sin\,\theta_{b}=0.5$. Here the solid (dashed) line represents
charged Higgs (total) contribution to the $b$ quark EDM without
extra dimension,  solid (dashed) curve represents charged Higgs
(total) single extra dimension contribution to the $b$ quark EDM
and dotted (double dashed) curve represents charged Higgs (total)
total contribution to the $b$ quark EDM. We see that the
contribution of the extra dimensions is larger than the one
without the extra dimension, for $1/R\leq 600\, GeV$. Here the
neutral Higgs contribution is suppressed similar to the top quark
EDM and we do not present in this figure.  The EDM of $b$ quark is
$5.0\times 10^{-20}\, (e-cm)$ for $1/R \sim 300\, GeV$ in the case
of the inclusion of a single extra dimension and it is more than
one order larger compared to the one without extra dimensions. It
is shown that the EDM of $b$ quark is sensitive to a single extra
dimension and this physical quantity is a candidate to test the
extra dimensions and to obtain stringent restrictions for the
compactification scale, with the accurate future experimental
measurements.

\begin{figure}[htb]
\vskip -3.0truein \centering \epsfxsize=6.8in
\leavevmode\epsffile{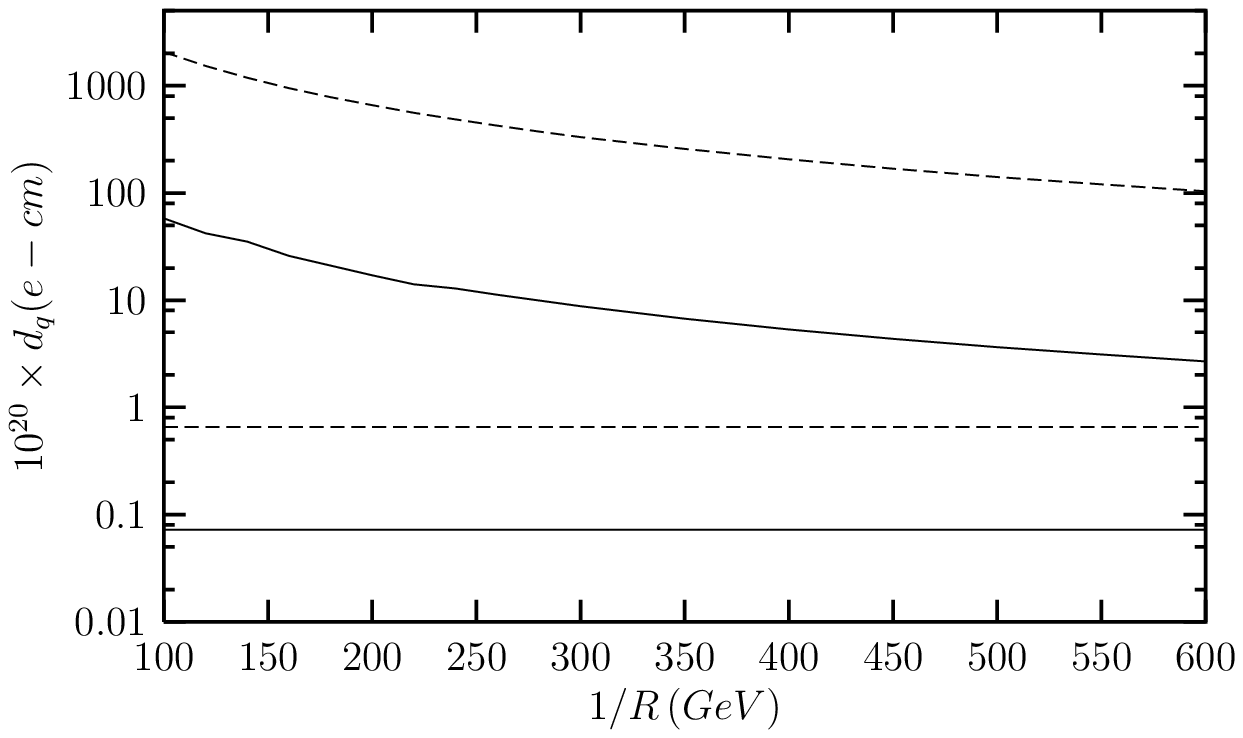} \vskip -3.0truein
\caption[]{The compactification scale $1/R$ dependence of $d_t$
and $d_b$  for $m_{H^{\pm}}=400\, GeV$, $m_{h^0}=85\, GeV$,
$m_{A^0}=90\, GeV$, and the intermediate value of
$sin\,\theta_{b}=0.5$ in the case of non-universal two extra
spatial dimensions. Here the solid (dashed) line represents $d_t$
($d_b$) without extra dimension, solid (dashed) curve represents
$d_t$ ($d_b$) with the inclusion of two extra dimensions.}
\label{EDMtbtotExtr2R}
\end{figure}

Fig. \ref{EDMtbtotExtr2R} shows the compactification scale $1/R$
dependence of $d_t$ and $d_b$ for $m_{H^{\pm}}=400\, GeV$,
$m_{h^0}=85\, GeV$, $m_{A^0}=90\, GeV$, and the intermediate value
of $sin\,\theta_{b}=0.5$ in the case of non-universal two extra
spatial dimensions. Here the solid (dashed) line represents $d_t$
($d_b$) without extra dimension,  solid (dashed) curve represents
$d_t$ ($d_b$) with the inclusion of two extra dimensions.It is
realized that the EDM of $t$ ($b$) quark is much more sensitive to
two extra dimensions and it is two orders (almost three orders)
larger compared to the one without extra dimensions, for $1/R\sim
300\, GeV$. This is a valuable information to test the
compactification scale and even the number of extra dimensions.

At this stage, we start to analyze the effects of the extra
dimensions on the charged lepton EDMs.  Here we make analysis for
$\mu$ and $\tau$ lepton and choose the numerical values
$\bar{\xi}^{E}_{N,\tau \mu} =10\, GeV$, $\bar{\xi}^{E}_{N,\tau
\tau}=50\, GeV$, $sin\theta_{\tau}=sin\theta_{\mu}=0.5$. The
assumption that there is no CKM type matrix in the leptonic sector
leads to the fact that there exist only neutral Higgs
contributions in the expression of EDM of charged leptons.

\begin{figure}[htb]
\vskip -3.0truein \centering \epsfxsize=6.8in
\leavevmode\epsffile{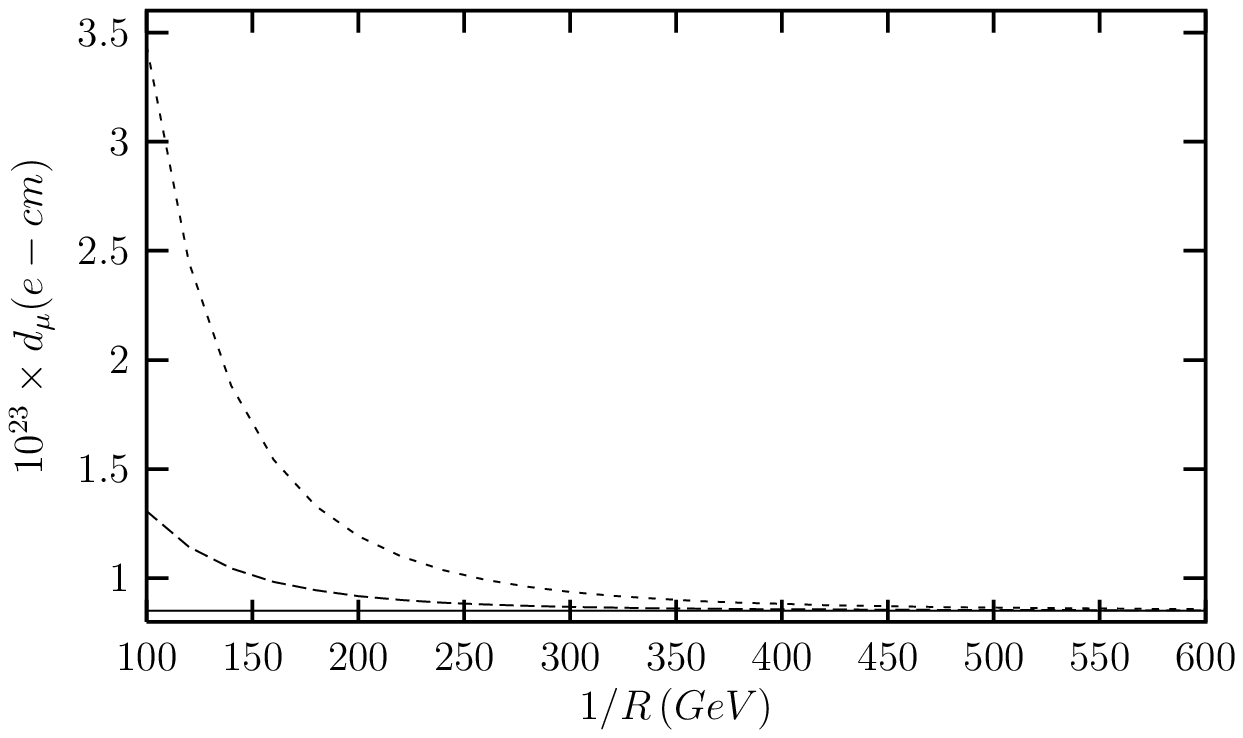} \vskip -3.0truein \caption[]{The
compactification scale $1/R$ dependence of $d_\mu$ for
$m_{h^0}=85\, GeV$, $m_{A^0}=90\, GeV$. Here the
solid-dashed-small dashed lines represent $d_\mu$ without-with a
single-with two extra dimensions.} \label{EDMmuR}
\end{figure}

\begin{figure}[htb]
\vskip -3.0truein \centering \epsfxsize=6.8in
\leavevmode\epsffile{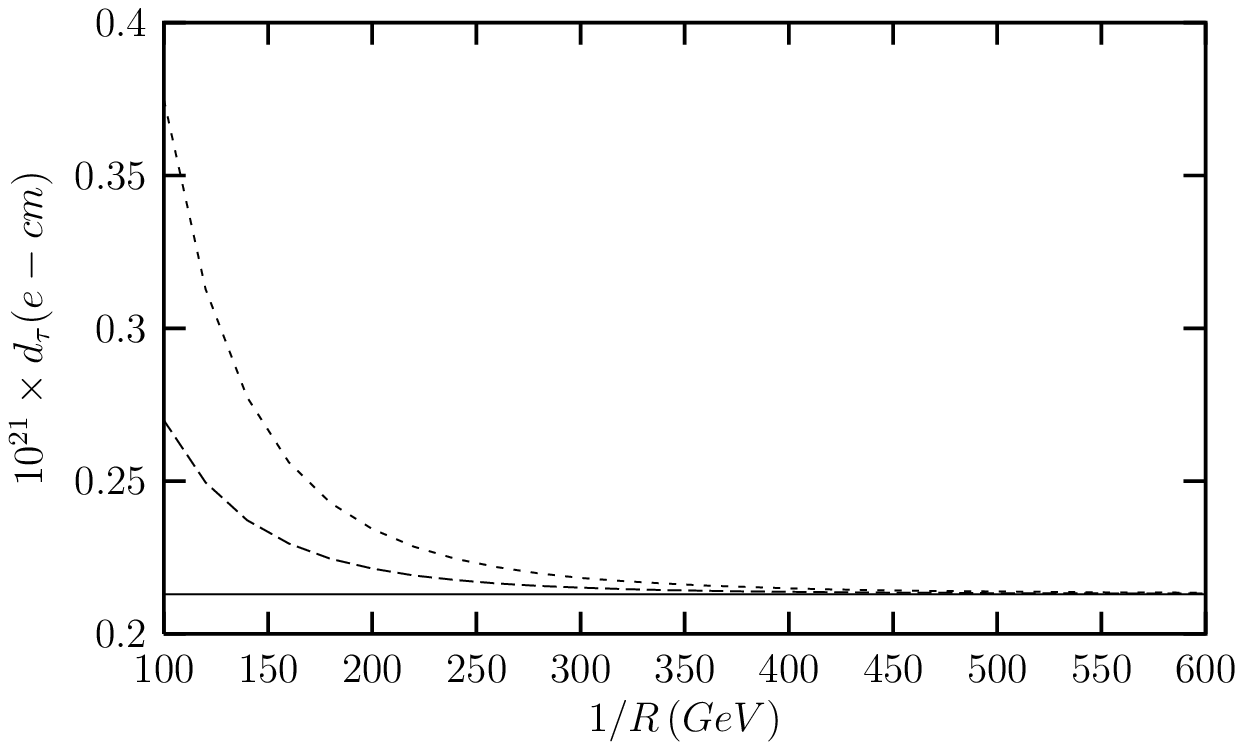} \vskip -3.0truein \caption[]{The
same as Fig. \ref{EDMmuR}, but for $d_\tau$.} \label{EDMtauR}
\end{figure}

In Fig. \ref{EDMmuR} (\ref{EDMtauR}) we present the
compactification scale $1/R$ dependence of $d_\mu$ ($d_\tau$) for
$m_{h^0}=85\, GeV$, $m_{A^0}=90\, GeV$. Here the
solid-dashed-small dashed  lines represent $d_\mu$ ($d_\tau$)
without-with a single-with two extra dimensions. It is observed
that $d_\mu$ ($d_\tau$) is weakly sensitive to the extra
dimensions for $1/R\geq 300\, GeV$ and the contribution due to the
two extra spatial dimension is almost $10\%$  for $1/R\sim 300\,
GeV$.

Now we would like to summarize our results:

\begin{itemize}

\item  The EDM of top quark is sensitive to the extra dimensions,
especially to two extra dimensions. The inclusion of a single
(double) extra dimension causes the EDM to enhance to the
numerical value $3.0\times 10^{-21}\, (e-cm)$ ($10^{-19}\,
(e-cm)$) for $1/R\sim 300\, GeV$ and it is almost six times (two
orders) larger than the one without extra dimensions. This is
informative to check the existence of extra dimensions and the
restriction of the compactification scale, with the help of the
forthcoming experimental measurements.

\item The EDM of $b$ quark is also sensitive to the extra
dimensions and it reaches the numerical value of $5.0\times
10^{-20}\, (e-cm)$ ($5.0\times 10^{-18}\, (e-cm)$) for $1/R\sim
300\, GeV$ in the case of the inclusion of a single (double) extra
dimension. The addition of extra dimensions results in the
enhancement of EDM of one order (three orders) compared to the one
without extra dimensions, for $1/R \sim 300\, GeV$. This is a
stong sensitivity and it can be used even to test the number of
extra dimensions, besides the restrictions for the
compactification scale.

\item  $d_\mu$ ($d_\tau$) is weakly sensitive to the extra
dimensions for $1/R\geq 300\, GeV$ and the contribution due to the
two extra spatial dimension is almost $10\%$  for $1/R\sim 300\,
GeV$.

\end{itemize}

Therefore, the experimental investigation of the EDMs of the top
quark and bottom quark can give powerful information about the
existence of extra dimensions.
\section{Acknowledgement}
This work has been supported by the Turkish Academy of Sciences in
the framework of the Young Scientist Award Program.
(EOI-TUBA-GEBIP/2001-1-8)
\newpage
\newpage
\newpage
\end{document}